\documentclass[aps,prd,twocolumn,superscriptaddress,nofootinbib]{revtex4-1}

\pdfoutput=1

\usepackage{graphicx}
\usepackage{amssymb,amsmath,latexsym}
\usepackage{xcolor}
\usepackage[normalem]{ulem}
\usepackage{subfigure}
\usepackage{placeins}
\usepackage{soul}
\usepackage[colorlinks=true,linktocpage=true,linkcolor=blue,citecolor=orange,urlcolor=blue]{hyperref}
\begin{document}
\title{Quantifying uncertainty in the neutron-star equation of state using point estimates and posterior distributions}
\author{Andr\'{e} G. da Silva} \email{andre-silva.as@acad.ufsm.br}
\affiliation{Departamento de F\'{i}sica, Universidade Federal de Santa Maria, 97105-900 Santa Maria, RS, Brazil}
\author{Ricardo L. S. Farias} \email{ricardo.farias@ufsm.br}
\affiliation{Departamento de F\'{i}sica, Universidade Federal de Santa Maria, 97105-900 Santa Maria, RS, Brazil}
\affiliation{Center for Nuclear Research, Department of Physics, Kent State University, Kent, OH 44242 USA}
%

%
%
\begin{abstract}

We investigate uncertainty quantification for the neutron-star equation of state (EOS) by comparing point-estimation and distributional inference approaches using the same Chebyshev and piecewise-linear parameterizations. We combine neutron-star mass--radius and gravitational-wave tidal-deformability information within Bayesian, multilayer-perceptron (MLP), and normalizing-flow frameworks. Although the methods yield similar mean EOS behavior, the deterministic MLP produces substantially narrower uncertainty bands at high densities. We show that this behavior is associated with the point-estimation objective, which maps degenerate solutions toward the conditional mean rather than representing the full parameter posterior. By contrast, the normalizing flow yields distributions more consistent with the Bayesian inference. Our results demonstrate that reliable uncertainty quantification of the high-density EOS requires methods that represent conditional probability distributions rather than only point estimates.

\end{abstract}
\maketitle

\section{Introduction}

The equation of state (EOS) of dense, strongly interacting matter is a central ingredient for understanding the internal composition and macroscopic properties of neutron stars (NSs). These stars can reach densities ranging from a few to approximately ten times nuclear saturation density and therefore are natural laboratories for studying quantum chromodynamics (QCD) at high densities. While QCD can be studied nonperturbatively at small baryon chemical potentials using lattice QCD (LQCD), the sign problem makes calculations at high densities exceptionally challenging~\cite{deForcrand:2009zkb,Nagata:2021ugx}. At baryon number densities $n_B \approx 40 n_s$ (where $n_s=0.16\text{fm}^{-3}$), asymptotic freedom implies that perturbative approaches to QCD (pQCD) lead to convergent results~\cite{Kurkela:2009gj,Kurkela:2016was,Gorda:2018gpy,Gorda:2021znl,Gorda:2023mkk}. Assuming that the high-density pQCD EOS can be connected consistently to the EOS realized inside NSs, it has been shown in Ref.~\cite{Komoltsev:2021jzg} that pQCD can provide valuable information about NS matter by imposing causality and thermodynamic stability, an approach subsequently applied in Refs.~\cite{Gorda:2022jvk,Komoltsev:2023zor,Gorda:2025aiu}. By contrast, at low temperatures and densities ($n_B \approx n_s$), chiral effective field theory ($\chi$EFT) provides a reliable expansion within which the equation of state and nuclear-matter properties can be calculated~\cite{Weinberg:1978kz,Epelbaum:2008ga,Drischler:2021kxf}. The intermediate-density region, however, is not fully constrained by pQCD and $\chi$EFT results alone.

The Tolman--Oppenheimer--Volkoff (TOV) equations~\cite{Tolman:1939jz,Oppenheimer:1939ne} provide a link between the equation of state and the macroscopic properties of stars and, as a result, allow the inference of the EOS by combining theoretical modeling with astrophysical observations such as masses, radii, and tidal deformabilities~\cite{Hinderer:2007mb,Flanagan:2007ix,Hinderer:2009ca}. A method to obtain the inverse map---from precise mass and radius observations to the EOS---was initially proposed in Ref.~\cite{Lindblom:1992}; however, to account for the ill-posed nature of the inverse problem and for observational uncertainties, statistical procedures must be used~\cite{Fujimoto:2021zas}. Subsequent works (see, e.g., Refs.~\cite{Steiner:2010fz,Ozel:2016oaf}) combined mass and radius measurements from quiescent low-mass X-ray binaries and thermonuclear bursts with a Bayesian inference framework to obtain constraints on the neutron-star equation of state. 

Since then, we have gained access to increasingly precise measurements~\cite{Riley:2019yda,Miller:2019cac,Riley:2021pdl,Miller:2021qha,Choudhury:2024xbk}, and new astrophysical messengers, such as gravitational waves~\cite{LIGOScientific:2017vwq}, have become available. In particular, mass--radius (MR) and mass--tidal-deformability (M$\Lambda$) observations are of great interest for inferring the EOS. Pulse-profile modeling can be used to estimate parameters such as the mass and equatorial radius from X-ray observations obtained with NICER~\cite{Riley:2019yda}. Meanwhile, the detection of gravitational waves from a binary neutron star merger has enabled the estimation of parameters such as the chirp mass $\mathcal{M}$ and the tidal parameter $\tilde\Lambda$, which enter the gravitational-wave signal observed by the LIGO/Virgo Collaboration~\cite{LIGOScientific:2017vwq,LIGOScientific:2018cki,LIGOScientific:2018hze}. Correspondingly, there has been increased interest in how multimessenger measurements can be combined and how they improve our knowledge of the NS EOS~\cite{Koehn:2024set}. 

Considerable attention has also been devoted to understanding the impact of different EOS parameterizations~\cite{Legred:2022pyp}, which can be divided into model-agnostic parameterizations and parameterizations based on effective models. Model-agnostic parameterizations do not assume a specific nuclear-matter model and are generally applied to study the relation between energy density and pressure, $p(\epsilon)$, with examples including the piecewise-polytrope, speed-of-sound, spectral-expansion~\cite{Lindblom:2010bb,Lindblom:2018rfr,Lindblom:2024yio,Greif:2018njt,Read:2008iy,Raithel:2016bux}. In addition to model-agnostic parameterizations, there are also nonparametric representations of the EOS~\cite{Landry:2018prl}. Conversely, parameterizations based on models allow the study of additional quantities that cannot be obtained from the relation $p(\epsilon)$ alone, such as the particle composition~\cite{Carvalho:2024kgf}. Some examples of the latter include relativistic mean-field and Nambu--Jona-Lasinio models~\cite{Carvalho:2023ele,Carvalho:2024kgf,Albino:2024ymc,Albino:2025puc,Marquez:2024bzj,Malik:2024qjw}. 

In recent years, Bayesian inference has become a standard framework for neutron-star EOS reconstruction, allowing for a systematic combination of observational data and explicit prior knowledge (for a non-exhaustive list, see Refs.~\cite{Landry:2018prl,Essick:2019ldf,Landry:2020vaw,Essick:2023fso,Miller:2019nzo,Raaijmakers:2019dks,Raaijmakers:2021uju,Albino:2024ymc,Albino:2025puc,Gorda:2022jvk,Annala:2023cwx,Komoltsev:2023zor,Koehn:2024set,Finch:2025bao,Brandes:2022nxa,Brandes:2023hma,Brandes:2024wpq,Legred:2021hdx,Mroczek:2023zxo,Altiparmak:2022bke,Marquez:2024bzj,daSilva:2023okq,Malik:2024qjw,Wouters:2025zju,Raithel:2017ity}). Within this approach, uncertainties in the inferred EOS are naturally quantified through the posterior probability distribution. Possible limitations of this framework include the definition of the likelihood function, which is usually constructed from posterior samples from MR and M$\Lambda$ analyses instead of directly from the underlying telescope or detector data~\cite{Farrell:2022lfd,Brandes:2024vhw}. Another limitation is the high computational cost associated with sampling posterior distributions, which is increased by the inclusion of more parameters. This computational cost also hinders the inclusion of new observations, since the entire posterior must generally be resampled whenever the likelihood is modified.

Parallel to these developments, neural-network methods in the context of simulation-based inference (SBI) have emerged as powerful tools for accelerating EOS inference and exploring high-dimensional parameter spaces. In particular, deterministic neural networks, such as multilayer perceptrons (MLPs), have been used to map neutron-star observables directly to EOS parameters or to predict macroscopic quantities from microscopic inputs (for a non-exhaustive list, see Refs.~\cite{Fujimoto:2017cdo,Fujimoto:2019hxv,Fujimoto:2021zas,Fujimoto:2024cyv,Soma:2022qnv,Soma:2022vbb,Morawski:2020izm,Chatterjee:2023ecc,Farrell:2022lfd,Ferreira:2019bny,Krastev:2023fnh,Zhou:2023cfs}). However, since a conventional deterministic MLP does not directly represent a posterior distribution, it is currently unclear whether such models can properly estimate the uncertainty of the inferred EOS. Another possible limitation of most implementations of SBI is that the models need to be retrained as the number of observations increases, although the training and inference steps might still be more computationally efficient than directly sampling the Bayesian posterior.

A comparison of some of the results obtained from MLPs with those obtained within a Bayesian framework, such as those obtained in Refs.~\cite{Fujimoto:2024cyv,Soma:2022vbb,Brandes:2022nxa,Landry:2020vaw}, shows that they agree, in general, on the mean behavior of the squared speed of sound. However, there is a discrepancy in the inferred high-density behavior of the squared speed of sound: while the Bayesian analysis predicts a large uncertainty, the neural-network analysis predicts a significantly smaller uncertainty. Other works employing a combination of Bayesian and neural-network methods obtain results that are more similar to the Bayesian approach~\cite{Carvalho:2023ele,Carvalho:2024kgf,Carvalho:2024zyb,Carvalho:2025qie,Ferreira:2024rnf,Thete:2022drz,Han:2021kjx,Han:2022sxt,Brandes:2024vhw,Morawski:2022aud}. However, since most analyses use different choices of parameterizations, priors, and observational datasets, a direct comparison among their results is difficult. 
Complementing the MLP approach, the posterior distribution of model parameters can be estimated using normalizing flows (NFs)~\cite{Papamakarios:2021,Kobyzev:2021nf}, which directly model probability distributions and provide distributional predictions\footnote{Normalizing flows are deterministic transformations for fixed inputs and latent variables. However, by transforming samples from a base distribution, they explicitly represent probability density functions and therefore provide distributional rather than point predictions.}. This technique has already been successfully used to accelerate gravitational-wave inference~\cite{Dax:2021tsq,Dax:2024mcn,Marx:2025ioo}, enabling real-time parameter estimation of merger events, which is crucial for the detection of electromagnetic counterparts.
By using the same parameterizations, namely, piecewise and Chebyshev-basis expansions, and following procedures that are as similar as possible for both the MLP and Bayesian approaches, we aim to compare the uncertainties in the inferred EOSs obtained by these methods. Additionally, we include an approach based on normalizing flows (also used in Refs.~\cite{Carvalho:2024zyb,Carvalho:2025qie,Brandes:2024vhw,Morawski:2022aud,Hu:2024lrj}) to elucidate the origin of the discrepancy in the resulting uncertainties between point and posterior-distribution estimation.

This paper is organized as follows. In Section~\ref{section2}, we present the microscopically stable and causal representations of the EOS used in this work. In Section~\ref{section3}, we develop the Bayesian inference framework, in particular, the definition of the likelihood function. In Section~\ref{section4}, we discuss the simulation-based inference approach implemented via multilayer perceptrons and normalizing flows. In Section~\ref{section5}, we discuss the EOS inference results for each parameterization and inference approach. Finally, in Section~\ref{section6}, we present our conclusions.

\section{EOS parameterizations}\label{section2}

\begin{figure*}[t]
    \centering
    \includegraphics[width=0.9\textwidth]{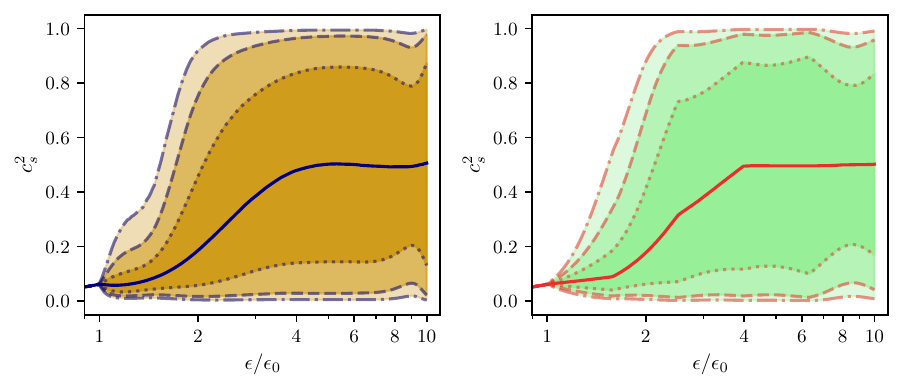}
    \caption{Speed of sound squared prior distributions generated for both the Chebyshev (left panel) and piecewise (right panel) bases. Solid lines show the median, dotted lines denote the $68\%$ CI, dashed lines denote the $95\%$ CI, and dash-dotted lines denote the $99\%$ CI.}
    \label{fig1}
\end{figure*}

The squared speed of sound, defined by $c_s^2 = dp/d\epsilon$, is an important quantity in modeling the EOS, as it describes the slope of the $p(\epsilon)$ curve. Microscopic equilibrium and causality imply that $0 \leq c_s^2 \leq 1$, a condition that needs to be considered when generating causal and stable EOSs. Instead, one can define an auxiliary quantity $\phi(\epsilon)$
\begin{equation}
    \phi(\epsilon) = \ln\left( \frac{1}{c_s^2(\epsilon)} - 1 \right),
\end{equation}
which takes values over the whole real line, naturally satisfying both causality and microscopic equilibrium~\cite{Lindblom:2010bb,Lindblom:2018rfr,Landry:2018prl}. Starting from $p_0 = p(\epsilon_0)$, we can integrate $c_s^2(\epsilon)$ to obtain the EOS:
\begin{equation}
    p(\epsilon) = p_0 + \int_{\epsilon_0}^{\epsilon} \frac{d\epsilon'}{1 + \exp[\phi(\epsilon')]}.
\end{equation}

Given that $\phi(\epsilon)$ is unconstrained, it is possible to define a spectral parameterization by expanding $\phi(\epsilon)$ in a set of basis functions. 
We consider two different expansions: one with parameters that affect the EOS over the entire density range and another with parameters that affect specific energy-density intervals. Chebyshev polynomials are often used as an expansion basis for functions defined in the interval $[-1,1]$. In this basis, $\phi(\epsilon)$ is expressed as~\cite{Lindblom:2024yio}
\begin{equation}
    \phi(\epsilon) = \sum_{i=0}^{N} \gamma_i T_i(z(\epsilon)),
\end{equation}
where $T_i(z)$ is the $i$-th Chebyshev polynomial of the first kind, $\gamma_i$ are the expansion parameters, and $z(\epsilon)$ is defined so that the polynomials are evaluated over the optimal interval (with $\epsilon_\text{max} = 10\epsilon_0$):
\begin{equation}
    z(\epsilon) = -1 + 2\ln\left( \frac{\epsilon}{\epsilon_0} \right) \left[ \ln \left( \frac{\epsilon_\text{max}}{\epsilon_0} \right) \right]^{-1}.
\end{equation}

To match the equation of state to a crust model at $\epsilon_0$, we must impose a condition on $\gamma_0$:
\begin{equation}
    \gamma_0 = \ln\left(\frac{1}{c_s^2(\epsilon_0)} - 1 \right) - \sum_{i=1}^{N} \gamma_i (-1)^i,
\end{equation}
where $c_s^2(\epsilon_0)$ and $p(\epsilon_0)$ are given by the crust EOS SKa~\cite{Kohler:1976fgx,Danielewicz:2008cm,Gulminelli:2015csa} (available in the CompOSE repository~\cite{Typel:2013rza,Oertel:2016bki,CompOSECoreTeam:2022ddl}), which is used throughout this paper.

The other parameterization considered here is also described in Ref.~\cite{Lindblom:2018rfr}, where $\phi(\epsilon)$ is expressed using piecewise segments that are linear in the logarithm of $\epsilon$ such that $\phi(\epsilon)$ takes the following form in the $k$-th segment:
\begin{equation}
    \phi(\epsilon) = \phi_k + \alpha_{k+1} \ln\left( \frac{\epsilon_k}{\epsilon_{k-1}} \right), \quad \epsilon_k \leq \epsilon \leq \epsilon_{k+1},
\end{equation}
where $\phi_k$ are constants constrained by continuity conditions, and $\alpha_{k+1}$ are the expansion parameters. Here, the energy density interval $[\epsilon_0, \epsilon_\text{max}]$ is divided into $5$ intervals equally spaced on a logarithmic scale, with $\epsilon_k$ denoting the interface between adjacent intervals. If we disregard correlations between parameters, each parameter in this piecewise parameterization affects only a single energy density interval.

However, generating random samples of either $\gamma_i$ or $\alpha_{k+1}$ from uniform distributions with the same span leads to highly biased distributions of $c_s^2$, as most samples will have $|\phi(\epsilon)|$ become very large. To remedy this, the priors on the parameters can be assigned using least-squares fits to $c_s^2(\epsilon)$ curves generated from a Gaussian process (GP). We follow a methodology adapted from Ref.~\cite{Gorda:2022jvk}, in which we change the independent variable from $n$ to $\epsilon$ and the GP mean hyperparameter distribution to a uniform distribution. We model $\phi(\epsilon)$ with a GP:
\begin{equation}
    \phi(\epsilon) \sim \mathcal{N}(-\ln(1/\bar{c}_s^2 - 1), K(\epsilon, \epsilon')),
\end{equation}
where we use the Gaussian kernel:
\begin{equation}
K(\epsilon, \epsilon') = \eta \exp(-(\epsilon-\epsilon')^2/2\ell^2.
\end{equation} 
The hyperparameters $\eta, \bar{c}_s^2$, and $\ell$ are sampled according to
\begin{equation}
\begin{aligned}
    \ell &\sim \mathcal{N}(1.0\epsilon_0, (0.25\epsilon_0)^2), \quad \eta \sim \mathcal{N}(1.25, 0.2^2), \\
    \bar{c}_s^2 &\sim \mathcal{U}(0.01, 0.99).
\end{aligned} 
\end{equation}

This GP is conditioned on two points: a low-noise constraint $(\mu_\phi = \phi_0, \sigma_\phi = 10^{-6})$ at $\epsilon = \epsilon_0$, and an intermediate-noise constraint $(\mu_\phi = -\ln(1/c_s^2 - 1), \sigma_\phi = 10^{-2})$ at $\epsilon = \epsilon_\text{max}$, where $c_s^2 \sim \mathcal{U}(0.01, 0.99)$. The last constraint was added to prevent a bias toward parameter sets that describe observational data. For example, observations might favor a stiffening at a particular density; however, EOSs in the prior that support this stiffening remain stiff at higher densities. 

Using both parameterizations and the least-squares fitting procedure described in Appendix~\ref{appendixA}, we generate prior samples of the $5$ independent parameters (note that $\gamma_0$ is a dependent parameter), resulting in the uniform-like distributions in $c_s^2(\epsilon)$ seen in Fig.~\ref{fig1}. The final prior distributions generated via the fitting process are approximated by multivariate Gaussian distributions, which are used in both the Bayesian and NN analyses. We use multivariate Gaussians as priors because the conversion to and from the unit cube (as required by many nested-sampling libraries) is well behaved and computationally inexpensive.

\section{Bayesian likelihoods}\label{section3}

Bayes' theorem allows us to relate the posterior probability distribution of the parameters of a model, $P(\boldsymbol{\theta} | \boldsymbol{D})$, to the likelihood of the observed data given the model parameters, $P(\boldsymbol{D} | \boldsymbol{\theta})$, and the prior distribution of these parameters, $P(\boldsymbol{\theta})$:
\begin{equation}
    P(\boldsymbol{\theta} | \boldsymbol{D}) = \frac{P(\boldsymbol{D} | \boldsymbol{\theta}) P(\boldsymbol{\theta})}{Z},
\end{equation}
where the normalization factor is the marginal likelihood obtained by integrating over the prior, called the evidence, which is typically a difficult high-dimensional integral. In the current setup, the parameters $\boldsymbol{\theta}$ are the parameters of the EOS parameterization under consideration, and the data $\boldsymbol{D}$ represent any directly measurable quantities, such as X-ray spectra or gravitational-wave signals. The latter are modeled via the likelihood, which is typically associated with aleatoric uncertainty, i.e., randomness in the data or irreducible noise. The posterior, on the other hand, is associated with epistemic uncertainty, i.e., uncertainty arising from limited data or from the model itself. 

\begin{figure*}[t]
    \centering
    \includegraphics[width=0.9\linewidth]{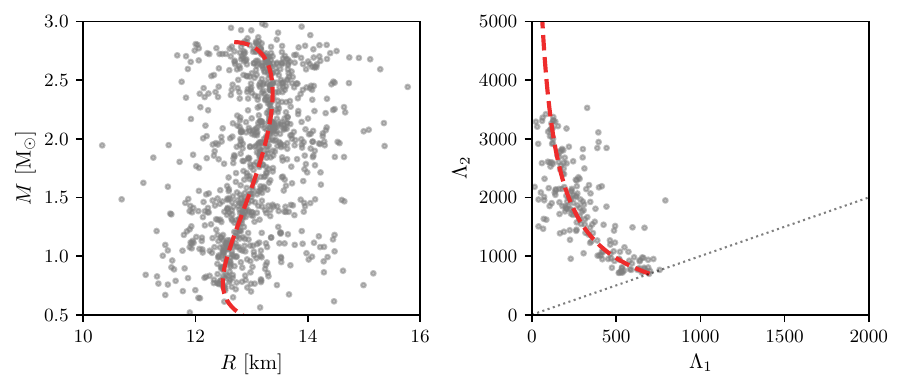}
    \caption{Generated $(M,R)$ (left panel) and $(\Lambda_1,\Lambda_2)$ (right panel) training and validation data. For visualization purposes, the right panel has a fixed chirp mass $\mathcal{M} = 1.186\text{M}_\odot$. The dotted line delimits the region where $\Lambda_1 > \Lambda_2$.}
    \label{fig2}
\end{figure*}

The observations are divided into two groups, referred to as NH and GW. The first group comprises mass-radius inferences for neutron stars obtained from NICER observations, together with the inference for the central compact object in the supernova remnant HESS J1731-347. Meanwhile, the second group comprises inferences of the masses and tidal deformabilities of a binary neutron star (BNS) system obtained from gravitational-wave observations by the LIGO/Virgo Collaboration. Specifically, we use the analyses performed in Refs.~\cite{Riley:2019yda,Riley:2021pdl} for the pulsars J0030+0451 and J0740+6620, which yield results that are qualitatively consistent with those obtained in Refs.~\cite{Miller:2019cac,Miller:2021qha}. Additionally, we also use the MR measurements of the millisecond pulsar J0437-4715~\cite{Choudhury:2024xbk} and those of the supernova remnant HESS J1731-347~\cite{Doroshenko:2022nwp}, which has been interpreted as being compatible with a neutron star, a hybrid star, or other more exotic compact-star configurations~\cite{Sagun:2023rzp}; however, its possible status as an outlier relative to the broader neutron-star population should be kept in mind~\cite{Koehn:2024set}. Considering, for simplicity, a single-branch mass-radius diagram, we can take the radius as a function of mass and formulate the likelihood as follows:
\begin{equation}
\begin{aligned}
    \mathcal{L}_\text{NH} = \prod_{i=1}^{N_\text{NH}} P(\boldsymbol{D}_i | \boldsymbol{\theta}) =  &\prod_{i=1}^{N_\text{NH}}\int P(\boldsymbol{D}_i | M, R, \boldsymbol{\theta}) 
    \\&\times P(M, R | \boldsymbol{\theta}) dM dR,
\end{aligned}
\end{equation}
where the product runs over all the MR observations.

If we assume that we can take the posterior samples from the NH analyses as observational data, it is possible to write the likelihood as
\begin{equation}
\begin{aligned} 
    P(\boldsymbol{D}_i | \boldsymbol{\theta}) \propto \int P(M, R_\theta(M) | \boldsymbol{D}_i, \boldsymbol{\theta}) P(M | \boldsymbol{\theta}) dM, 
\end{aligned}
\end{equation}
where we have used Bayes' theorem,
\begin{equation}
P(\boldsymbol{D}_i | M, R, \boldsymbol{\theta}) = \frac{P(M, R | \boldsymbol{D}_i, \boldsymbol{\theta}) P(\boldsymbol{D}_i | \boldsymbol{\theta})}{P(M, R | \boldsymbol{\theta})},
\end{equation}
and the fact that the TOV equations provide, for a single-branch family, a deterministic link between mass and radius, given the EOS parameters:
\begin{equation}
    P(R | M, \boldsymbol{\theta}) = \delta(R - R_\theta(M)).
\end{equation}
Here, we make the assumption that the NH analysis priors are sufficiently flat, such that the weights $w = 1/P(M,R_\theta(M) | \boldsymbol{\theta})$ contribute only a constant to the likelihood~\cite{Brandes:2022nxa}.

Then, we take the EOS-dependent mass prior to be uniform and adopt the following form for the mass priors
\begin{equation}
    P(M | \boldsymbol{\theta}) = \left\{ 
    \begin{array}{cl} 
        \frac{1}{M_\text{max}(\boldsymbol{\theta})- M_\text{min}} & \text{if} \quad M_\text{min} \leq M \leq M_\text{max}(\boldsymbol{\theta}), \\ 
        0 & \text{otherwise},
    \end{array} 
    \right.
\end{equation}
where $M_\text{min} = 0.7\text{M}_\odot$ and $M_\text{max}(\boldsymbol{\theta})$ is the maximum mass supported by the EOS. The inverse dependence on $M_\text{max}(\boldsymbol{\theta})$ acts as an Occam factor, favoring EOSs that produce maximum masses that are not much larger than the highest observed mass~\cite{Landry:2020vaw}. Although the mass prior should depend on the population model, such that lighter NSs do not pull the MR posterior toward lower masses, we decided to use only one mass prior for all observations to better match the SBI results, which are expected to carry an Occam factor through the mass sampling procedure.

This form of the likelihood can be better understood if we take a Gaussian approximation for $P(M, R_\theta(M))$ and consider the maximization of $\ln \mathcal{L}_{\text{NH}}$ for a single observation, given by
\begin{equation}
    \ln \mathcal{L}_{\text{NH}} \propto -\sum_i \frac{(R_\theta(M_i) - \mu_R)^2}{2\sigma_R^2} + \text{const.},
\end{equation}
where $\mu_R$ is the mean radius and $\sigma_R$ is the standard deviation of the radius. The maximization of such a likelihood with respect to $\boldsymbol{\theta}$ is equivalent to minimizing the mean squared error (MSE). The choice not to use central pressures or energy densities allows the Bayesian model to scale better with the number of observations, since otherwise the number of parameters would increase with each new MR measurement.

We also use the posterior samples for the masses and tidal deformabilities of the binary neutron star (BNS) merger GW170817, observed by the LIGO/Virgo Collaboration~\cite{LIGOScientific:2017vwq,LIGOScientific:2018hze,LIGOScientific:2018cki}. Another BNS merger event, GW190425, was detected by the LIGO/Virgo Collaboration~\cite{LIGOScientific:2020aai}; however, the posterior obtained for the component tidal deformabilities is not restrictive enough to be used in this work. The likelihood for the GW170817 observation is then given by
\begin{equation}
\begin{aligned}
    \mathcal{L}_\text{GW} = \int &P(q, \Lambda_1(M_1), \Lambda_2(M_2) | \boldsymbol{D}, \boldsymbol{\theta}, \mathcal{M}) \\
    &\times P(q | \boldsymbol{\theta}) dq,
\end{aligned}
\end{equation}

Following Ref.~\cite{Raaijmakers:2019dks}, we fix the chirp mass $\mathcal{M}~=~1.186 \text{M}_\odot$\footnote{The convention $M_1 > M_2$ is used throughout this paper.}, since it is well constrained and allows for better density estimation. The prior $P(q | \boldsymbol{\theta})$ is taken to be a uniform distribution between $0.5$ and $1.0$.

The final likelihood is given by the product of the two contributions:
\begin{equation}
    \mathcal{L} = \mathcal{L}_\text{NH} \times \mathcal{L}_\text{GW}.
\end{equation}

In the current procedure, the densities $P(M,R_\theta(M) | \boldsymbol{D}_i,\boldsymbol{\theta})$ and $P(q, \Lambda_1(M_1), \Lambda_2(M_2) | \boldsymbol{D}, \boldsymbol{\theta}, \mathcal{M})$ need to be evaluated within the integral. These densities can be estimated from samples of the observables in two ways: via a Gaussian approximation or via a kernel density estimate (KDE). In the first procedure, the mean vector and covariance matrix of the observables are estimated from the samples, giving rise to CIs that correspond to ellipses. In the KDE procedure, the density is estimated nonparametrically using a kernel function and a smoothing matrix. The KDE procedure captures more complex shapes, such as those present in the MR and M$\Lambda$ posteriors. However, it is currently unclear whether accounting for non-Gaussian noise leads to qualitatively different results from those obtained using the Gaussian approximation.

\section{Simulation-based inference}\label{section4}

Simulation-based inference (SBI), or likelihood-free inference, is a technique used when the likelihood function has a high computational cost or when likelihood-density estimation suffers from high-dimensionality problems~\cite{Papamakarios:2021,Cranmer:2019eaq}. This is achieved by simulating how samples from the likelihood are generated, e.g., by sampling observations predicted by a deterministic model and adding noise to them. In particular, we use a form of SBI called neural posterior estimation (NPE), in which the NN model is trained to directly approximate the posterior distribution of the EOS parameters (see, e.g., Ref.~\cite{Carvalho:2025qie}). Alternatively, the model may instead approximate the likelihood, an approach known as neural likelihood estimation (NLE) (see, e.g., Ref.~\cite{Brandes:2024vhw}).

In the context of inferring the neutron star EOS using a parametric model, we draw samples from the prior distribution and solve the TOV equations to obtain the curves $M(p_c)$, $R(p_c)$, and $\Lambda(p_c)$, where $p_c$ is the pressure at the center of the star. In agreement with the definition of the Bayesian likelihood, we assume that there is a single stable branch for which we can obtain both $R$ and $\Lambda$ as functions of $M$. We then sample $N_\text{NH}$ points from the uniform distribution in mass $\mathcal{U}(0.7\text{M}_\odot, M_\text{max})$, obtaining the pairs $(M_i, R(M_i))$. In order to sample $(\Lambda_1,\Lambda_2)$ pairs, which are parameters inferred in the GW analysis, we proceed by fixing the chirp mass $\mathcal{M}$. For this purpose, we sample the chirp mass from the uniform distribution $\mathcal{U}(1.17\text{M}_\odot, 1.20\text{M}_\odot)$ and obtain the two curves $\Lambda_1(M_1)$ and $\Lambda_2(M_2)$. A mass ratio $q_i$ is sampled from the uniform distribution $\mathcal{U}(0.5, 1.0)$, which, together with the chirp mass, determines a pair $(\Lambda_{1,i}, \Lambda_{2,i})$. The sampling of pairs is performed $n_s=20$ times and, for each pair, we resample $n_g=10$ points according to the following prescription:
\begin{equation}
\begin{aligned}
    &M \sim \mathcal{N}(M_i, \sigma_{M}^2), &R \sim \mathcal{N}(R_i, \sigma_{R}^2), \\
    &\sigma_{M} \sim \mathcal{U}(0, 0.2\text{M}_\odot), &\sigma_{R} \sim \mathcal{U}(0, 1.2\text{km}), \\
    &\Lambda_1 \sim \mathcal{N}(\Lambda_{1,i}, \sigma_{\Lambda_1}^2), &\Lambda_2 \sim \mathcal{N}(\Lambda_{2,i}, \sigma_{\Lambda_2}^2), \\
    &\sigma_{\Lambda_1} \sim \mathcal{U}(0, 240), &\sigma_{\Lambda_2} \sim \mathcal{U}(0, 365), \\
    &q \sim \mathcal{N}(q_i, \sigma_q^2), &\sigma_q \sim \mathcal{U}(0, 0.1).
\end{aligned}
\end{equation}
The maximum values of the standard deviations are chosen so that they are consistent with the observations used in this work. This method is similar to those used in Refs.~\cite{Fujimoto:2024cyv,Carvalho:2025qie}, with some modifications to include a fixed chirp mass.

Although SBI is often called likelihood-free inference, it only avoids the explicit evaluation of the likelihood function~\cite{Cranmer:2019eaq}. The simulator model, which describes how the parameters lead to noisy observations, nevertheless implicitly defines a likelihood for each observation. In the preceding discussion, this likelihood is similar to an uncorrelated Gaussian distribution for MR measurements. For this reason, when comparing the Bayesian and SBI results, we use only the Gaussian likelihood results. 

The final result of the generation of the training data for a particular EOS can be seen in Fig.~\ref{fig2}. We generate a total of $100\text{,}000$ EOSs from the multivariate Gaussian distributions of the fitted parameters discussed in Sec.~\ref{section2}, replacing those that fail to support stars with masses of at least $0.7\text{M}_\odot$\footnote{Such procedure modifies the prior, but is also present in the Bayesian likelihood calculation.}. Then, we split the data into $90\%$ training and $10\%$ validation sets, which are used to train both the multilayer perceptron and normalizing-flow models.

\subsection{Multilayer perceptron}\label{section4A}

Feedforward neural networks, or multilayer perceptrons (MLPs), are machine-learning methods inspired by the functioning of biological neurons. Universal approximation theorems provide a theoretical foundation showing that MLPs can approximate broad classes of functions~\cite{Goodfellow:2016dl}. Each hidden layer in the MLP uses parameterized affine transformations composed with nonlinear functions, progressively constructing a representation of the target function. In this work, we refer to the NN inputs as input features, the hidden neurons as hidden features, and the network outputs as output features. The output of the $j$-th unit in the $l$-th layer is given by
\begin{equation}
    a_j^l = \sigma\left( \sum_k w_{jk}^l a_k^{l-1} + b_j^l \right),
\end{equation}
where $w_{jk}^l$ and $b_j^l$ are trainable parameters, and $\sigma$ is the activation function. 

Given the MLP architecture, we can define a loss function that measures the discrepancy between the predicted outputs and the ground truth, and minimize it to determine the parameter values that best fit the training data. A convenient loss for regression problems is the mean squared error (MSE), given by
\begin{equation}
    L(\boldsymbol{\phi}) = \frac{1}{N} \sum_{i=1}^N \lVert \boldsymbol{\hat{y}}(\boldsymbol{x}_i; \boldsymbol{\phi}) - \boldsymbol{y}_i \rVert^2,
\end{equation}
where $N$ is the number of training samples and $\boldsymbol{x}_i$ denotes the $i$-th training data sample. The NN output $\boldsymbol{\hat y}(\boldsymbol{x}_i, \boldsymbol{\phi})$ is given by the composition of each NN layer and depends on the network parameters $\boldsymbol{\phi}$. Then, the predicted output is compared with the expected output $\boldsymbol{y}$ (often called the label). It can be shown that an NN that minimizes this loss function estimates the conditional mean~\cite[see ch.~4]{Bishop:2023}:
\begin{equation}
    \hat{\boldsymbol{y}}(\boldsymbol{\phi}) = \mathbb{E}[\boldsymbol{y} | \boldsymbol{x}],
\end{equation}
and, within this context, different loss functions lead to different estimated statistics of the parameter posterior. The mean squared logarithmic error, for example, leads to a form close to the conditional geometric mean. Apart from aleatoric uncertainty (i.e., uncertainty that is irreducible even in the limit of infinite data), an MLP trained with the MSE loss function will estimate only a single statistic of the target, the conditional mean. 

Even though MLPs are defined to make point predictions, Monte Carlo (MC) integration techniques can be employed to calculate expectation values and estimate the sensitivity of inferred parameters to observational uncertainty. The expectation value of the EOS parameters $\boldsymbol{\theta}$ under the observational distribution $P(\boldsymbol{D})$ is given by~\cite{Fujimoto:2024cyv}
\begin{equation}
    \mathbb{E}_{P(D)}[\boldsymbol{\theta}] = \int \hat{\boldsymbol{y}}(\boldsymbol{D}; \boldsymbol{\phi}) P(\boldsymbol{D}) d\boldsymbol{D},
\end{equation}
where $\boldsymbol{x}$ denotes the NN inputs. Using an MC estimate with $N$ samples drawn from the distribution $p(\boldsymbol{x})$, the integral can be approximated as\footnote{A similar procedure can be applied to quantiles, such as the median and equal-tail credible intervals.}
\begin{equation}
    \mathbb{E}_{P(D)} [\boldsymbol{\theta}] = \frac{1}{N} \sum_{i=1}^N \hat{\boldsymbol{y}}(\boldsymbol{D}_i; \boldsymbol{\phi}).
\end{equation}

\begin{figure*}[t]
    \centering
    \includegraphics[width=0.7\linewidth]{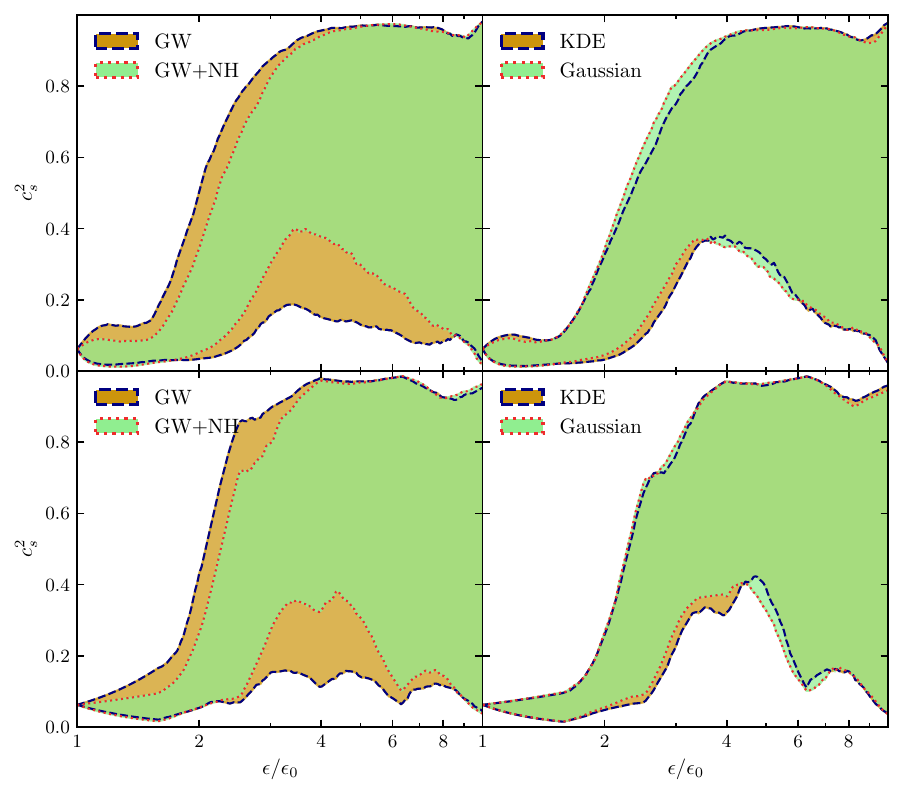}
    \caption{Squared speed of sound posterior CIs for the Chebyshev (top panels) and piecewise linear (bottom panels) parameterizations. In the left panels, dashed lines show the $68\%$ CI of the posterior obtained using only the GW likelihood, while dotted lines show the $68\%$ CI of the posterior obtained using the combined GW+NH likelihood. In the right panels, dashed lines show the $68\%$ CI of the posterior obtained using KDEs as density estimators for the GW+NH analysis posteriors, while dotted lines show the $68\%$ CI of the posterior obtained using Gaussian fits as density estimators.}
    \label{fig3}
\end{figure*}

\subsection{Normalizing flow}\label{section4B}

Normalizing flows (NF) provide a framework in which density estimation, that is, estimating the density or probability distribution of samples, can be achieved through continuous and invertible transformations. Starting from a base distribution $P_u(\boldsymbol{u})$, we express the transformed variable $\boldsymbol{x}$ as~\cite{Papamakarios:2021}
\begin{equation}
    \boldsymbol{x} = T(\boldsymbol{u}),
\end{equation}
where $\boldsymbol{u}$ is sampled from the base distribution and $T$ is the flow transformation, which can be a composition of several simpler transformations. The probability density $P_x(\boldsymbol{x})$ is obtained using the change-of-variables formula~\cite{Papamakarios:2021}:
\begin{equation}
\begin{aligned}
    P_x(\boldsymbol{x}) &= P_u(\boldsymbol{u}) \left| \det J_T(\boldsymbol{u}) \right|^{-1}, \\
    (J_T)_{ij} &= \frac{\partial T_i}{\partial u_j}, 
\end{aligned}
\end{equation}
where the factor $|\det J_T(\boldsymbol{u})|^{-1}$ guarantees that $P_x(\boldsymbol{x})$ remains a valid probability density function. The process of obtaining the distribution of $\boldsymbol{x}$ from the distribution of $\boldsymbol{u}$ is denoted the generative direction.

Given enough transformations composing $T$, an NF can approximate a broad class of non-trivial continuous probability distributions starting from a simpler base distribution. This follows from the fact that, like MLPs, normalizing flows also have universality properties~\cite{Kobyzev:2021nf}. It is possible to parameterize $T$ (e.g., with residual NN blocks) and train an NF to approximate the required transformation. This requires computing the inverse transformation $T^{-1}$ together with its Jacobian determinant, a process denoted the normalizing direction.

The goal of finding a flow that transforms $P_u(\boldsymbol{u})$ into $P_\theta(\boldsymbol{\theta} | \boldsymbol{D})$ is achieved by minimizing the expected Kullback--Leibler (KL) divergence between the target distribution $P_\theta^*(\boldsymbol{\theta} | \boldsymbol{D})$ and the one obtained by the NF, $P_\theta(\boldsymbol{\theta} | \boldsymbol{D}; \boldsymbol{\phi})$. We can express the loss function (up to a constant in $\boldsymbol{\phi}$) as~\cite{Papamakarios:2021,Carvalho:2025qie}
\begin{equation}
\begin{aligned}
    L(\boldsymbol{\phi}) &= \mathbb{E}_{P(D)} \big[ D_\text{KL}\big(P_\theta^*(\boldsymbol{\theta} | \boldsymbol{D}) \, \| \, P_\theta(\boldsymbol{\theta} | \boldsymbol{D}; \boldsymbol{\phi})\big)\big], \\
    &= \int (d\boldsymbol{D}) P(\boldsymbol{D}) \int d\boldsymbol{\theta} P_\theta^*(\boldsymbol{\theta} | \boldsymbol{D}) \ln\left( \frac{P_\theta^*(\boldsymbol{\theta} | \boldsymbol{D})}{P_\theta(\boldsymbol{\theta} | \boldsymbol{D} ; \boldsymbol{\phi})} \right),
\end{aligned}
\end{equation}
and, by identifying $P_\theta(\boldsymbol{\theta} | \boldsymbol{D} ; \boldsymbol{\phi})$ as the distribution obtained through the generative direction and applying MC integration, we arrive at
\begin{equation}
\begin{aligned}
    L(\boldsymbol{\phi}) = &-\frac{1}{M N} \sum_{i=1}^{M} \sum_{j=1}^{N} \bigg[ \ln P_u (T^{-1}(\boldsymbol{\theta}_i | \boldsymbol{D}_j ; \boldsymbol{\phi})) \\ 
    &+ \log \left| \det J_{T^{-1}}(\boldsymbol{\theta}_i | \boldsymbol{D}_j; \boldsymbol{\phi}) \right| \bigg] + \text{const.},
\end{aligned}
\end{equation}
where $M$ and $N$ are the numbers of parameter and context samples, respectively. Given context samples $\boldsymbol{D}_j$, the posterior distribution of the parameters can be estimated using the generative direction.

\begin{figure*}[p]
    \centering
    \includegraphics[width=1.0\linewidth]{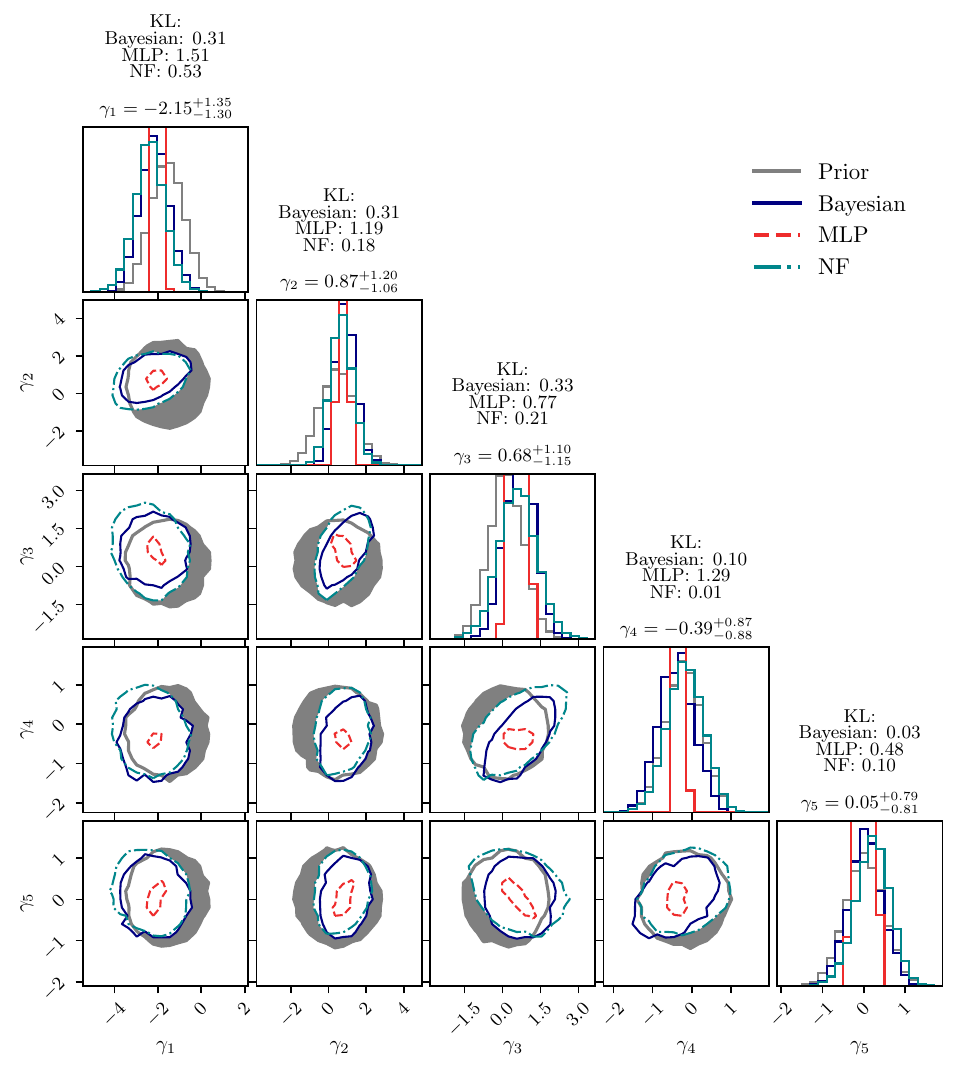}
    \caption{Posterior marginals of the Chebyshev basis parameters. The two-dimensional marginals show the $95\%$ CIs obtained by the Bayesian (solid lines), MLP (dashed lines), and NF (dash-dotted lines) approaches. The gray shaded regions show the $95\%$ CIs of the prior. At the top of each one-dimensional marginal, the KL divergence is given in units of nats for each ML approach. The parameter values shown correspond to the median and $95\%$ CI of the Bayesian posterior.}
    \label{fig4}
\end{figure*}

\begin{figure*}[p]
    \centering
    \includegraphics[width=1.0\linewidth]{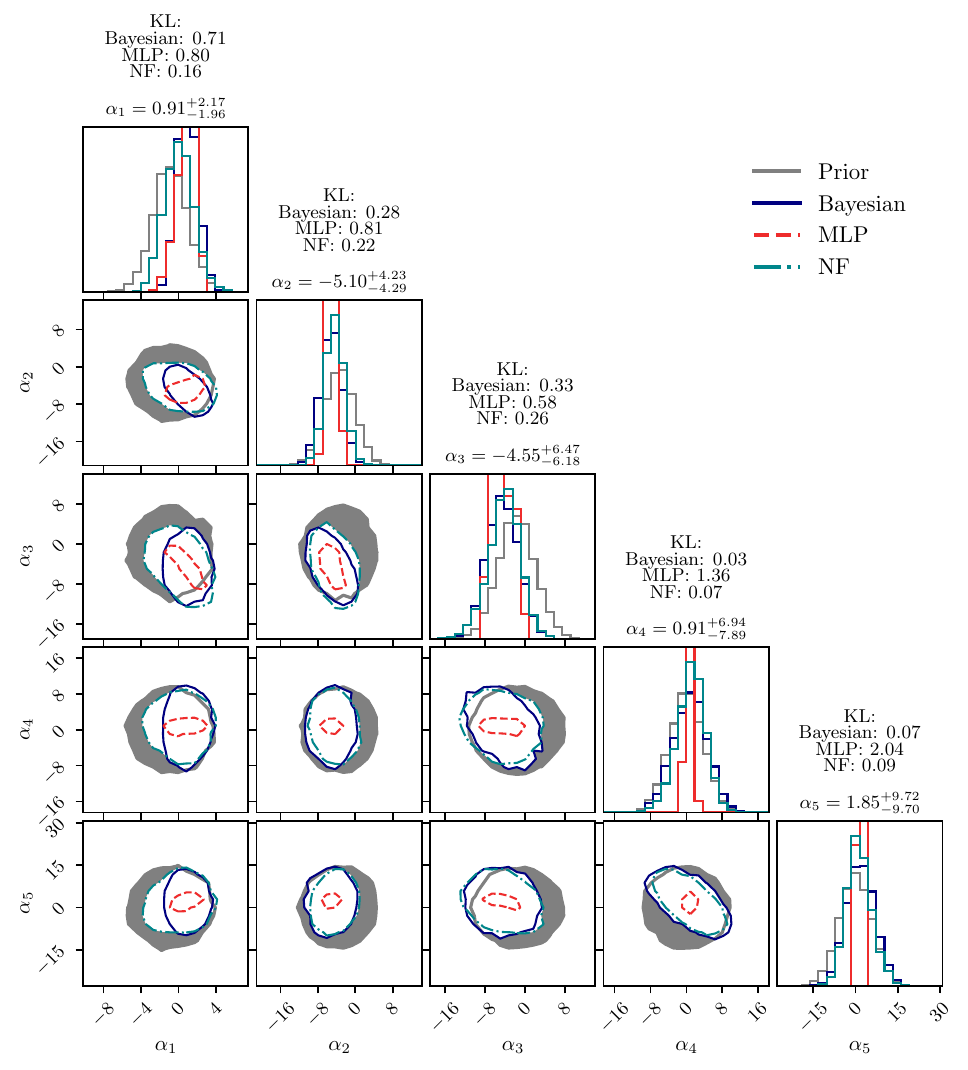}
    \caption{Posterior marginals of the piecewise linear parameters. The two-dimensional marginals show the $95\%$ CIs obtained by the Bayesian (solid lines), MLP (dashed lines), and NF (dash-dotted lines) approaches. The gray shaded regions show the $95\%$ CIs of the prior. At the top of each one-dimensional marginal, the KL divergence is given in units of nats for each ML approach. The parameter values shown correspond to the median and $95\%$ CI of the Bayesian posterior.}
    \label{fig5}
\end{figure*}

\begin{figure*}[t]
    \centering
    \includegraphics[width=0.95\linewidth]{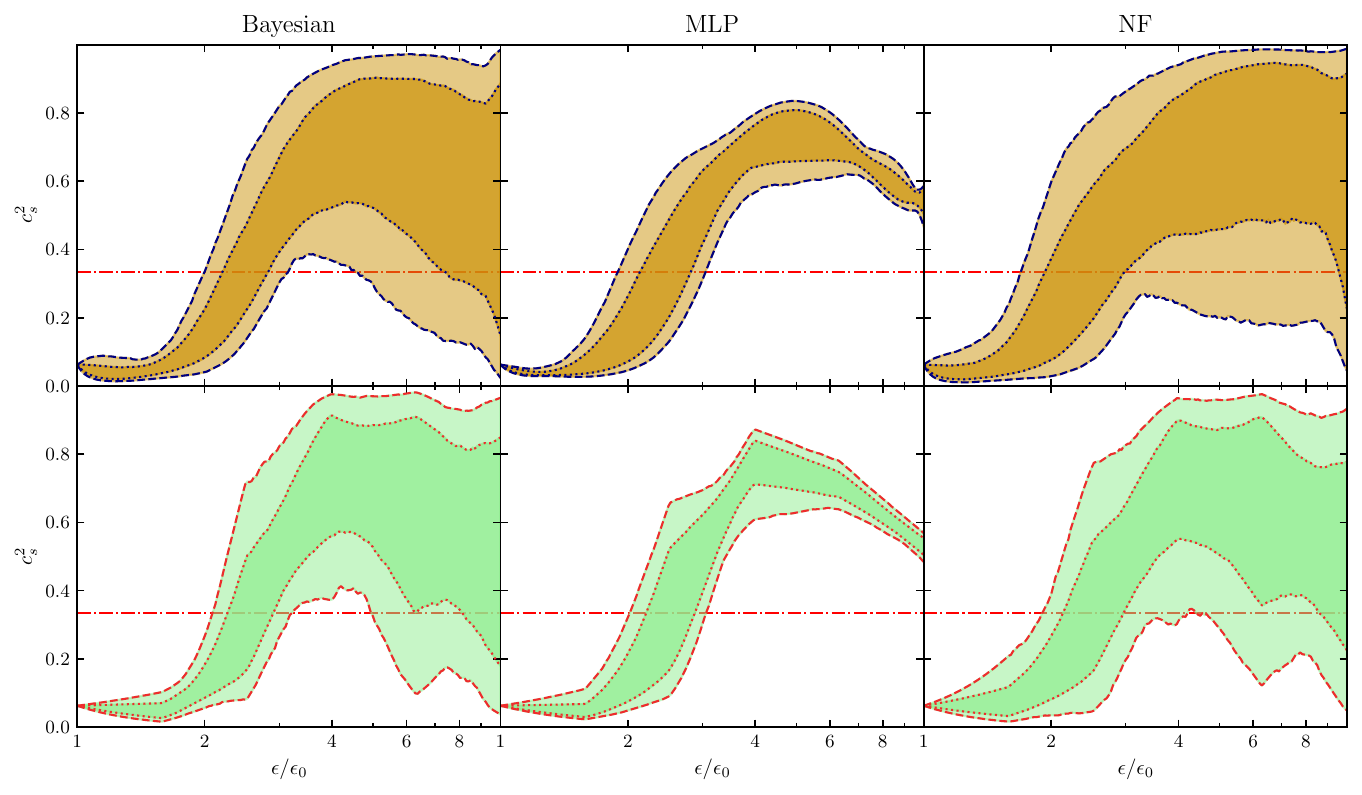}
    \caption{Squared speed of sound posterior CIs for the Chebyshev (top panels) and piecewise linear (bottom panels) parameterizations. Dotted lines and dashed lines indicate the $68\%$ and $95\%$ levels, respectively. The horizontal dash-dotted lines indicate the conformal limit, $c_s^2 = 1/3$. The left column corresponds to the Bayesian approach, the middle column to the MLP approach, and the right column to the NF approach.}
    \label{fig6}
\end{figure*}

 \section{EOS inference results}\label{section5}

\subsection{Impact of observations and likelihood density estimators}\label{section5A}

With the implementation details described in Appendix~\ref{appendixB}, we perform Bayesian nested sampling for the two parameterizations. First, we investigate the constraining capability of each type of observation. The left panels of Fig.~\ref{fig3} show the results obtained using only the GW likelihood, compared with those obtained using the combined GW$+$NH likelihood. The observational constraints favor a stiffening of the EOS at densities $\epsilon \approx 3\epsilon_0$, with the combined likelihood imposing tighter constraints. Next, we test the validity of the likelihood constructed with a Gaussian density estimator relative to the likelihood constructed with a KDE, as shown in the right panels of Fig.~\ref{fig3}. It appears that the Gaussian likelihood leads to EOSs that become stiffer at slightly smaller densities, although the effect is mostly negligible. While the first result is also important as a sanity check of the NH and GW likelihoods, the second result justifies the use of Gaussian likelihoods for comparison with the SBI-based approaches, since they rely on Gaussian sampling of the calculated MR distributions.

\subsection{Inferences from different machine-learning approaches}\label{section5B}

Having established a baseline using the Bayesian inference approach, we proceed to perform SBI with the two NN models, with implementation details described in Appendix~\ref{appendixB}. We present the parameter posterior\footnote{In the MLP approach these are the prediction distributions, which quantify the sensitivity of the inferred parameters to observational uncertainties.} marginals for the Chebyshev basis parameters, shown in Fig.~\ref{fig4}, where the Python library \texttt{corner}~\cite{corner} was used. The corner plot shows that both the Bayesian and NF approaches lead to similar uncertainties in the parameters, while the MLP model results in lower overall uncertainty. Since the MLP and NF share the same SBI framework, the difference between the two methods arises from their different objectives. This is evident in the marginal posteriors, which show an apparent collapse of the MLP predictions toward the mean of the NF distributions. Correlations between pairs of parameters appear as tilted credible regions. It was found that the Bayesian and NF posteriors show a positive correlation between adjacent parameters and a smaller correlation between nonadjacent parameters, however, such correlations are not easily interpretable. Since the KL divergence is a measure of the dissimilarity between two distributions, it also enables us to quantify how much information was gained about a specific parameter. Here, all models show different KL divergence values across all parameters, although it needs to be emphasized that the single-parameter KL divergence does not include correlations between parameters.  

We also present the parameter posterior marginals for the piecewise-linear parameters, shown in Fig.~\ref{fig5}, for the piecewise-linear parameterization. The plot shows that both the Bayesian and NF approaches lead to similar posteriors for most parameters, except for the first two parameters. Most importantly, the lower uncertainty attained by the MLP approach is visible by inspecting both the one- and two-dimensional marginal posteriors of the EOS parameters. Here, the collapse of the MLP predictions to the conditional mean can also be visualized. Turning our attention to the KL divergence, it was found that, while both the Bayesian and NF results show greater information gain at lower energy densities and little information gain at higher densities, the MLP inference results in the largest gain across all energy-density intervals and, specifically, in the highest-density intervals. This behavior is also seen in Fig.~\ref{fig5} and indicates a possible collapse of the MLP results toward a specific set of EOS parameters. It can be seen that, although the prior parameters are relatively uncorrelated, the parameter posteriors of the Bayesian and NF approaches show correlations between neighboring intervals, i.e., in the diagonal two-dimensional marginal posteriors. Unlike the correlations found in the Chebyshev parameters, the correlation between adjacent parameters is easily interpretable: an increase in a slope parameter must be followed by a decrease in the next slope parameter in order for the resulting EOS to remain approximately unchanged, thereby producing a negative correlation between the two parameters. It can be seen that such correlations might depend on the specific parameterization, which illustrates the parameterization dependence inherent in parametric approaches~\cite{Legred:2022pyp}.

Next, we turn our attention to the posterior of the squared speed of sound, which can be visualized in Fig.~\ref{fig6} for both Chebyshev-based and piecewise-linear parameterizations. We found that both parameterizations lead to similar mean behavior for the same machine-learning setup. Although the Chebyshev parameters have similar uncertainties and the piecewise parameters show larger uncertainties at higher densities, the final $c_s^2$ posteriors of the two parameterizations are more alike. Comparing the results obtained with the different ML approaches, we found that, at densities $\epsilon \approx 3\epsilon_0$, all approaches indicate a stiffening of the EOS in order to describe the observational constraints. At higher densities, $\epsilon > 8\epsilon_0$, all setups show median $c_s^2$ values of approximately $0.5$, which indicates that the posterior is largely dominated by the prior. In this region, the MLP uncertainties differ drastically from the Bayesian and NF uncertainties, attaining low uncertainty in a possibly unconstrained region, as indicated by the small information gain and similarity between the prior and posterior distributions. One interpretation is that, when different values of the parameters lead to predictions that cannot be distinguished given the observational uncertainty, point-wise predictors will collapse all output values to the predicted statistic. The difference between the two NN models lies in their objectives: the MLP approach is designed to infer the conditional mean of the parameters, whereas the NF approach is designed to approximate the posterior distribution.

\begin{figure*}[t]
    \centering
    \includegraphics[width=0.95\linewidth]{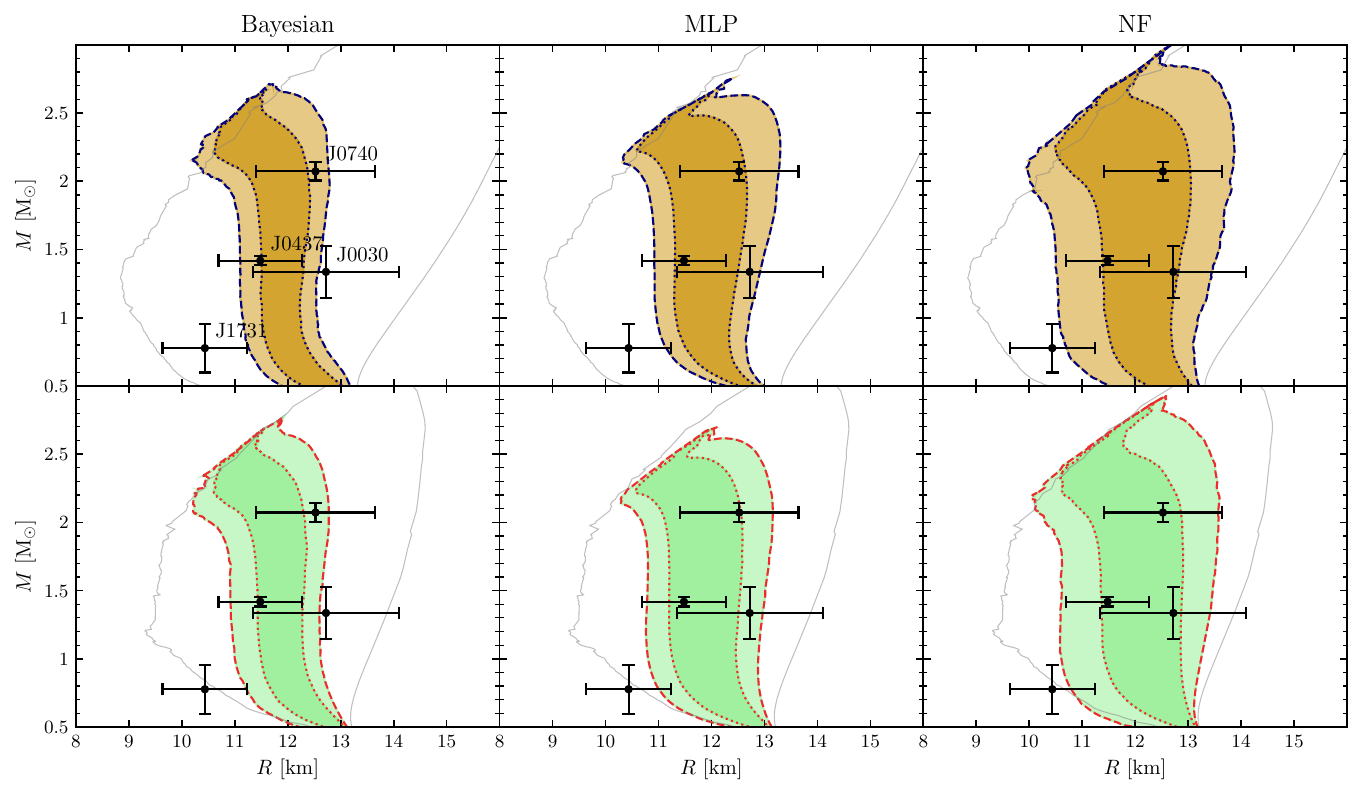}
    \caption{Mass-radius diagram posterior CIs for the Chebyshev (top panels) and piecewise linear (bottom panels) parameterizations. Dotted lines and dashed lines indicate the $68\%$ and $95\%$ levels, respectively. The black error bars indicate the NS measurements used in the analysis, fitted with Gaussian distributions. The left column corresponds to the Bayesian approach, the middle column corresponds to the MLP approach, and the right column corresponds to the NF approach. Gray lines show the $95\%$ CIs of the priors.}
    \label{fig7}
\end{figure*}

\begin{figure*}
    \centering
    \includegraphics[width=0.9\linewidth]{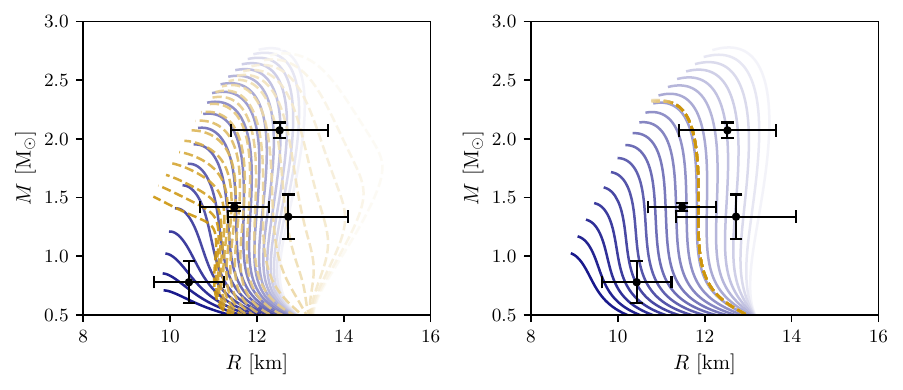}
    \caption{Mass-radius curves obtained by varying the first (solid lines) and last (dashed lines) EOS parameters, for both the Chebyshev (left panel) and piecewise-linear (right panel) parameterizations. The black error bars indicate the NS measurements used in the analysis, fitted with Gaussian distributions. Solid lines show variations in the first parameters, while dashed lines show variations in the last parameters, which range from lower (more transparent) to higher (more opaque) values.}
    \label{fig8}
\end{figure*}

Using samples from the parameter posteriors, we calculate the MR diagram posteriors for all approaches and parameterizations, which can be seen in Fig.~\ref{fig7}. It can be seen that, although the MLP posteriors have much lower uncertainties in the speed of sound than the other approaches, they still lead to similar uncertainties in the observables. This is a result of the nonlinear nature of the mapping defined by the TOV equations, in which large uncertainties in the EOS are converted into smaller uncertainties in the NS observables. To reinforce this idea, we take the Bayesian parameter posterior samples and fix all parameters but one, which is varied uniformly between its lowest and highest attained values. The result of this process can be seen in Fig.~\ref{fig8}, where varying the first and last parameters leads to distinguishable MR curves relative to the observational errors in the Chebyshev parameterization, while only the first parameter generates distinguishable curves in the piecewise-linear parameterization. In the Chebyshev parameterization, the uncertainty in the high-density regime might lie in the correlation between parameters, since, as also shown in Fig.~\ref{fig4}, all parameters have similar uncertainties. Taking advantage of the fact that the $\alpha_{k+1}$ parameters each apply only to specific energy-density intervals, this result, combined with the calculated KL divergence, reinforces the idea that the neutron star EOS is less constrained at higher energy-density intervals by current observations. It should be emphasized that, while there might be stable NSs that reach such high densities, the high-density EOS may still have little impact on the mass, radius, and tidal deformability of such stars. Within this context, even though future observations that aim to reduce radius uncertainties to a few hundred meters~\cite{STROBE-XScienceWorkingGroup:2019cyd,eXTP:2016rzs,Watts:2018iom,Li:2025uaw,Chatziioannou:2021tdi,Huxford:2023qne} will better constrain the EOS at intermediate densities, the highest densities might remain unconstrained still. 

\section{Conclusions}\label{section6}

In this work, we have inferred constraints on the equation of state of dense matter using three complementary inference approaches. We have used a combination of mass and radius posteriors from NICER observations and the analysis of the compact object in HESS J1731--347 with mass and tidal-deformability posteriors from the LIGO/Virgo detectors. Our analysis focused on the EOS parameters, the squared speed of sound, and mass-radius posteriors. We also employed two parametric forms of the EOS, one containing global parameters and the other containing parameters that, when correlations are neglected, control individual energy-density segments, in order to study how observations impact the intermediate- and high-density regions of the EOS.

In the Bayesian approach, we obtain the EOS posteriors using either the GW measurements alone or the combined GW and NH measurements. It was found that, while GW measurements constrain the $c_s^2$ posterior, the combined observations yielded much tighter constraints. We have also checked that Gaussian approximations to the observational probability densities entering the likelihood lead to posteriors similar to those obtained using KDEs. This comparison was performed to facilitate a more consistent comparison with the SBI methods, since the latter make use of implicit likelihoods via the simulator model.

We have found that the MLP model predicts much lower uncertainties in the high-density region, compared with the other two models. This is attributed to the lack of information about the high-density EOS in the observables, as quantified by the KL divergence. We also emphasize that an MLP trained with an MSE loss function predicts the conditional mean of the parameters $\mathbb{E}[\boldsymbol{\theta}|\boldsymbol{D}]$, which is only a single statistic of the posterior. Conversely, the normalizing-flow model predicts posteriors that are more closely aligned with the Bayesian results, which is attributed to its explicit modeling of the full posterior distribution. Other neural-network approaches can also provide probabilistic inference, including Bayesian neural networks, neural likelihood estimation, and probabilistic neural-network parameterizations of the EOS. What these approaches have in common is that they aim to represent probability distributions rather than only point predictions, which is essential for reliable uncertainty quantification. 

Overall, the present observations favor a stiffening of the EOS at intermediate densities, while the EOS at the highest densities remain only weakly constrained. Our results also demonstrate that approaches aimed at modeling distributions are essential for avoiding artificially narrow uncertainty estimates in this regime.

\bigskip \noindent {\it Acknowledgments.} This work was partially supported by Conselho Nacional de Desenvolvimento Cient\'ifico e Tecno\-l\'o\-gico  (CNPq), Grants No. 312032/2023-4, 402963/2024-5 and 445182/2024-5 (R.L.S.F.); Funda\c{c}\~ao de Amparo \`a Pesquisa do Estado do Rio 
Grande do Sul (FAPERGS), Grant No. 24/2551-0001285-0 (R.L.S.F.); CAPES Finance Code 001 (A.G.S); The work is also part of the project
Instituto Nacional de Ci\^encia e Tecnologia - F\'isica Nuclear e
Aplica\c{c}\~oes (INCT - FNA), Grants No. 464898/2014-5 and 408419/2024-5, and supported
by the Ser\-ra\-pi\-lhei\-ra Institute (grant number Serra -
2211-42230). R. L. S. F. acknowledges
the kind hospitality of the Center for Nuclear Research at Kent State University, where part of this work was
done.


\bigskip \noindent {\it Data Availability.} The data are not publicly available. The data are available from the authors upon reasonable request.

\bibliographystyle{apsrev4-2}
\bibliography{refs}

\appendix

\section{Linear least-squares fit of $\phi(\epsilon)$}\label{appendixA}

With the expansion of $\phi(\epsilon)$ using the Chebyshev basis, $\phi(\epsilon) = \sum_i \gamma_i T_i(z(\epsilon))$, we define a grid based on the GP samples, denoted by $(z(\epsilon_g),\phi_g)$. Combining this expansion with the crust-matching condition, we arrive at the equation
\begin{equation}
    \phi(z(\epsilon_g)) - \phi_0 = \sum_{i=0}^{N} \gamma_i \left[ T_i(z(\epsilon_g)) - (-1)^ i \right],
\end{equation}
where, by comparison with the linear equation $\boldsymbol{b}~=~A \boldsymbol{\gamma}$, we identify $b_g = \phi(z(\epsilon_g)) - \phi_0$ (defined over the grid space), $\boldsymbol{\gamma} = \gamma_i$ (defined on parameter space), and $A_{gi}~=~\left[ T_i(z(\epsilon_g)) - (-1)^ i \right]$ (defined over both the grid and parameter spaces), where $i~=~0,1,2,\dots,N$.

\begin{figure*}[t]
    \centering
    \includegraphics[width=0.9\linewidth]{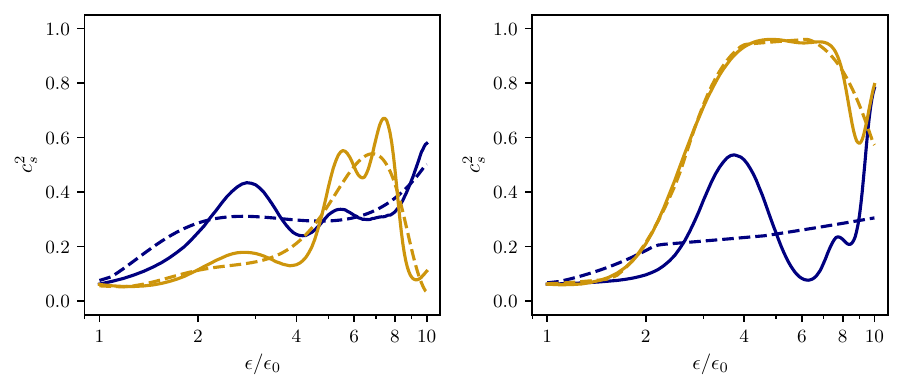}
    \caption{Example of generated EOSs using a Chebyshev-basis fit (left panel) and a piecewise-linear fit (right panel) for $2$ samples from the GP prior. Solid lines show the non-parametric GP samples, while dashed lines show the basis fits using three (light yellow curves) and five (dark blue curves) independent parameters.}
    \label{fig9}
\end{figure*}

The same procedure can be applied to the piecewise-linear expansion, where the crust-matching condition gives
\begin{equation}
    \phi_m = \phi_{m-1} + \alpha_m \ln \left( \frac{\epsilon_m}{\epsilon_{m-1}} \right),
\end{equation}
leading to
\begin{equation}
\begin{aligned}
    \phi(\epsilon_g) - \phi_0 = \sum_{k=1}^{m-1}\alpha_k \ln \left( \frac{\epsilon_k}{\epsilon_{k-1}} \right) + \alpha_m \ln\left( \frac{\epsilon_g}{\epsilon_{m-1}} \right), \\ \quad \epsilon_{m-1} \leq \epsilon_g \leq \epsilon_m.
\end{aligned}
\end{equation}
We can again write this expression in the form of a linear equation:
\begin{equation}
    \phi(\epsilon_g) - \phi_0 = \sum_{k=1}^{N}\alpha_k C_{g k},
\end{equation}
where the design matrix $C_{jk}$ is given by
\begin{equation}
    C_{gk} = \begin{cases}
        \ln\left( \frac{\epsilon_k}{\epsilon_{k-1}} \right) \quad &\text{if} \quad k < m, \\
        \ln\left( \frac{\epsilon_g}{\epsilon_{m-1}} \right) \quad &\text{if} \quad k = m, \\
        0 \quad &\text{if} \quad k > m.
    \end{cases}
\end{equation}

Figure~\ref{fig9} shows the resulting basis-function fits to the GP samples of $\phi(\epsilon)$ after transformation to $c_s^2(\epsilon)$. The figure shows that increasing the number of parameters improves the quality of the fits (i.e., leading to lower residuals).

\section{Machine-learning implementation}\label{appendixB}

\subsection{Bayesian posterior samples}

To define the prior distribution, we make use of $100\text{,}000$ EOS samples generated according to the prescription detailed in Appendix~\ref{appendixA} and calculate a multivariate Gaussian approximation to the basis parameters obtained from the least-squares fit. This approximation simplifies the transformation to the unit cube. Then, the \texttt{UltraNest}~\cite{Buchner:2021} nested-sampling Python library is used, which internally relies on the MLFriends~\cite{Buchner:2019} algorithm to propose new samples. We generate posterior samples using a minimum of $700$ live points and a slice sampler with $30$ steps per iteration, which incorporates Markov chain Monte Carlo within the nested-sampling algorithm to improve the proposed samples.

\subsection{MLP architecture and optimization}

We define an MLP with three hidden layers, each with $64$ hidden features and a \texttt{ReLU} activation function. The input features are composed of the $(M,R)$ pairs for the observed neutron stars together with $(\mathcal{M},q,\Lambda_1,\Lambda_2)$, resulting in a total of $12$ features, while the output features are the five independent EOS parameters. Both the input $\boldsymbol{x}$ and the output $\boldsymbol{y}$ are normalized to have zero mean and unit standard deviation:
\begin{equation}
    \boldsymbol{x}' = \frac{\boldsymbol{x} - \boldsymbol{\mu}_x}{\boldsymbol{\sigma}_x}, \quad \boldsymbol{y}' = \frac{\boldsymbol{y} - \boldsymbol{\mu}_y}{\boldsymbol{\sigma}_y},
\end{equation}
where $\boldsymbol{\mu}_{x}$, $\boldsymbol{\mu}_{y}$, $\boldsymbol{\sigma}_{x}$, and $\boldsymbol{\sigma}_{y}$, are the means and standard deviations of each component of the input and output, with vector division operations applied element-wise. Finally, we perform the training with the Adam~\cite{Kingma:2017} optimizer using a mini-batch size of $10,000$ and a learning rate of $0.001$. 

\subsection{NF architecture and optimization}

The normalizing-flow architecture is based on the masked autoregressive flow (MAF) architecture~\cite{Papamakarios:2017maf}, which models the conditional distribution $P(\boldsymbol{\theta} | \boldsymbol{x}_\text{obs})$, where $\boldsymbol{\theta}$ are the EOS parameters and $\boldsymbol{x}_\text{obs}$ are the NS observables, that is, $(M,R)$ pairs and $(\mathcal{M}, q, \Lambda_1, \Lambda_2)$. The model is composed of $16$ transformation layers, each consisting of a reverse permutation and a masked affine autoregressive transformation, which internally uses three residual neural-network blocks, each with $128$ hidden features. We use the Python library \texttt{nflows}~\cite{Durkan:2020nflows} to implement this flow architecture and the Adam optimizer to train the NF with a mini-batch size of $10,000$ and a learning rate of $0.0001$. The input parameters and context variables are normalized to have zero mean and unit standard deviation.
\end{document}